\documentclass[conference]{IEEEtran}
\IEEEoverridecommandlockouts
\usepackage{cite}
\usepackage{amsmath,amssymb,amsfonts}
\usepackage{algorithmic}
\usepackage{graphicx}
\usepackage{textcomp}
\usepackage{xcolor}
\def\BibTeX{{\rm B\kern-.05em{\sc i\kern-.025em b}\kern-.08em
    T\kern-.1667em\lower.7ex\hbox{E}\kern-.125emX}}

\usepackage{soul}

\newenvironment{myquote}%
  {\list{}{\leftmargin=0in\rightmargin=0in}\item[]}%
  {\endlist}

\usepackage{etoolbox}
\usepackage{caption}

\newcommand{\insertfig}{\setcounter{figure}{0}
\includegraphics[width=\linewidth]
{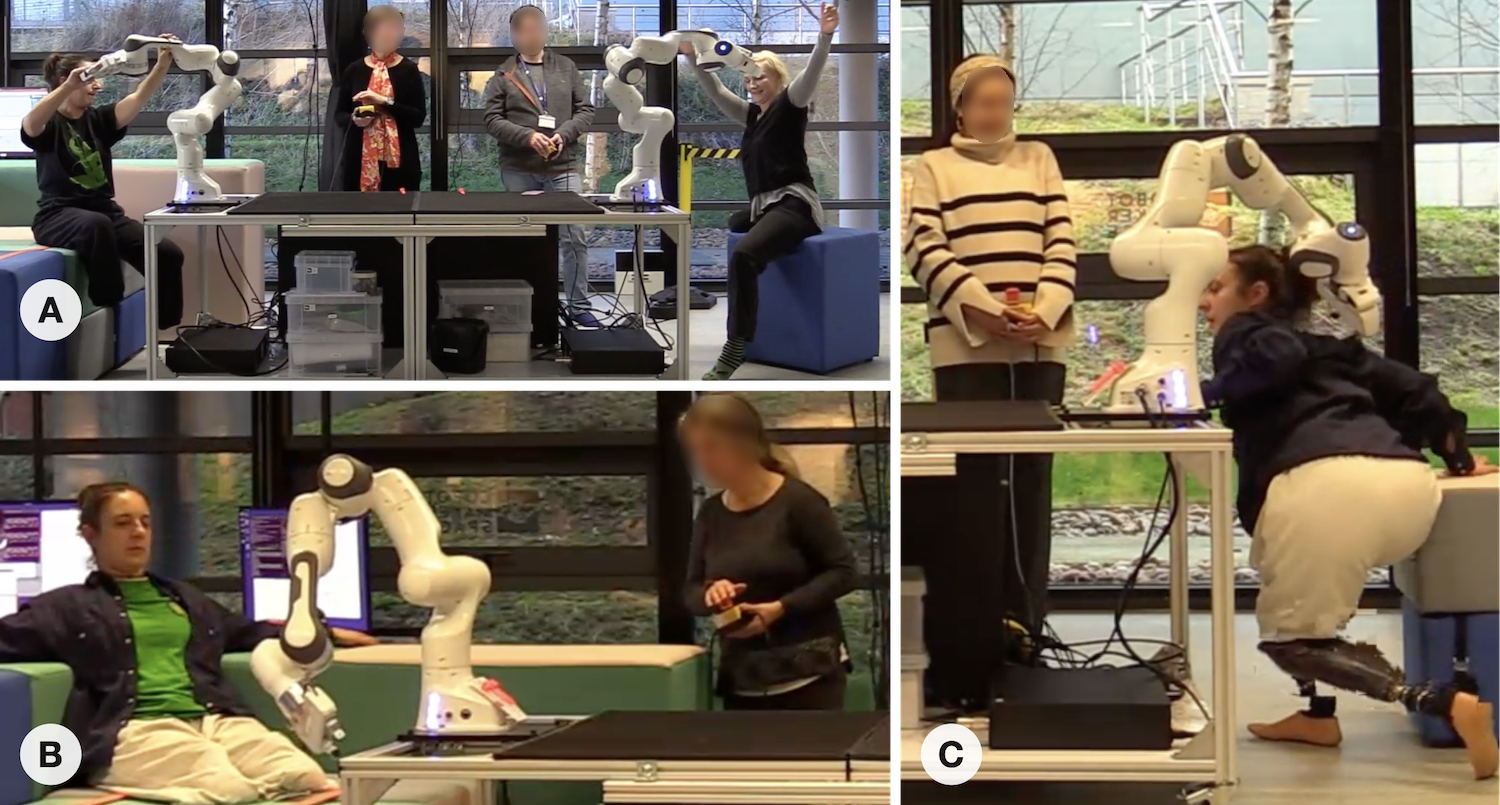}\captionof{figure}{\textit{\textbf{A:}} A dancer uses both hands to wiggle a robot moments before it freezes. \textit{\textbf{B:}} The wrong prerecorded movement replays on a robot, causing it to move surprisingly close to the table and leading the operator to activate the emergency stop. \textit{\textbf{C:}} A dancer closely brushes their back against a robot while the emergency stop operator looks on uncomfortably.}\label{fig:teaserfig}}

\makeatletter
\apptocmd{\@maketitle}{\centering\insertfig}{}{}
\makeatother

\usepackage{tcolorbox}
\newtcolorbox{mybox}[1]{%
    tikznode boxed title,
    enhanced,
    arc=0mm,
    interior style={white},
    attach boxed title to top center= {yshift=-\tcboxedtitleheight/2},
    fonttitle=\bfseries,
    colbacktitle=white,coltitle=black,
    boxed title style={size=normal,colframe=white,boxrule=0pt},
    title={#1}}

\usepackage{float}
\usepackage{caption}
\usepackage{capt-of}
\DeclareCaptionType{vignette}[Transcript][List of Vignettes]

\begin{document}

\title{Somatic Safety: An Embodied Approach Towards Safe Human-Robot Interaction\\
}

\author{
\IEEEauthorblockN{Steve Benford}
\IEEEauthorblockA{\textit{School of Computer Science} \\
\textit{University of Nottingham}\\
Nottingham, United Kingdom\\
steve.benford@nottingham.ac.uk}
\and
\IEEEauthorblockN{Eike Schneiders}
\IEEEauthorblockA{\textit{School of Electronics and Computer Science} \\
\textit{University of Southampton}\\
Southampton, United Kingdom \\
eike.schneiders@soton.ac.uk}
\and
\IEEEauthorblockN{Juan Pablo Martinez Avila}
\IEEEauthorblockA{\textit{School of Computer Science} \\
\textit{University of Nottingham}\\
Nottingham, United Kingdom\\
j.avila@nottingham.ac.uk}
\and
\IEEEauthorblockN{Praminda Caleb-Solly}
\IEEEauthorblockA{\textit{School of Computer Science} \\
\textit{University of Nottingham}\\
Nottingham, United Kingdom\\
praminda.caleb-solly@nottingham.ac.uk}
\and
\IEEEauthorblockN{Patrick Robert Brundell}
\IEEEauthorblockA{\textit{School of Computer Science} \\
\textit{University of Nottingham}\\
Nottingham, United Kingdom\\
pat.brundell@nottingham.ac.uk}
\and
\IEEEauthorblockN{Simon Castle-Green}
\IEEEauthorblockA{\textit{School of Computer Science} \\
\textit{University of Nottingham}\\
Nottingham, United Kingdom\\
simon.castle-green@nottingham.ac.uk}
\and
\IEEEauthorblockN{Feng Zhou}
\IEEEauthorblockA{\textit{School of Computer Science} \\
\textit{University of Nottingham}\\
Nottingham, United Kingdom\\
feng.zhou2@nottingham.ac.uk}
\and
\IEEEauthorblockN{Rachael Garrett}
\IEEEauthorblockA{\textit{Media Technology and Interaction Design} \\
\textit{KTH Royal Institute of Technology}\\
Stockholm, Sweden\\
rachaelg@kth.se}
\and
\IEEEauthorblockN{Kristina Höök}
\IEEEauthorblockA{\textit{Media Technology and Interaction Design} \\
\textit{KTH Royal Institute of Technology}\\
Stockholm, Sweden\\
khook@kth.se}
\and
\IEEEauthorblockN{Sarah Whatley}
\IEEEauthorblockA{\textit{Research Centre for Dance Research} \\
\textit{Coventry University}\\
Coventry, United Kingdom\\
adx943@coventry.ac.uk}
\and
\IEEEauthorblockN{Kate Marsh}
\IEEEauthorblockA{\textit{Research Centre for Dance Research} \\
\textit{Coventry University}\\
Coventry, United Kingdom\\
ac4988@coventry.ac.uk}
\and
\IEEEauthorblockN{Paul Tennent}
\IEEEauthorblockA{\textit{School of Computer Science} \\
\textit{University of Nottingham}\\
Nottingham, United Kingdom\\
paul.tennent@nottingham.ac.uki}
}

\IEEEoverridecommandlockouts
\IEEEpubid{\makebox[\columnwidth]{978-1-5386-5541-2/18/\$31.00~\copyright2018 IEEE \hfill} \hspace{\columnsep}\makebox[\columnwidth]{ }}

\maketitle

\IEEEpubidadjcol

\begin{abstract}
As robots enter the messy human world so the vital matter of safety takes on a fresh complexion with physical contact becoming inevitable and even desirable. We report on an artistic-exploration of how dancers, working as part of a multidisciplinary team, engaged in contact improvisation exercises to explore the opportunities and challenges of dancing with cobots. We reveal how they employed their honed bodily senses and physical skills to engage with the robots aesthetically and yet safely, interleaving improvised physical manipulations with reflections to grow their knowledge of how the robots behaved and felt. We introduce somatic safety, a holistic mind-body approach in which safety is learned, felt and enacted through bodily contact with robots in addition to being reasoned about. We conclude that robots need to be better designed for people to hold them and might recognise tacit safety cues among people.We propose that safety should be learned through iterative bodily experience interleaved with reflection.
\end{abstract}

\begin{IEEEkeywords}
Robotics, human-robot interaction, safety, dance, somatic safety, soma design
\end{IEEEkeywords}

\section{Introduction}
As robots spread into the messy everyday human world, physical contact with humans becomes inevitable and, at times, even desirable. Social and domestic applications from cleaning \cite{fink2013living} to care giving \cite{vallor2020carebots} to dancing \cite{Rogel:2022:Dancing} involve physical interaction with robots at close quarters. This naturally raises the important matter of safety.

While there has been extensive research into safe human-robot interactions within HRI \cite{lasota2017survey}, this has been dominated by reasoning about safety ‘at a distance’, by which we mean addressing safety concerns by humans carefully  ‘thinking them through’ prior to contact, for example through safety assessments, regulation, and programming robots to avoid or mitigate collisions 
\cite{zacharaki2020safety}. We explore the proposition that \textit{reasoning} about safety, while certainly necessary, is not sufficient. We adopt the stance that safety is also inherently a bodily concern; involving our pre-conscious actions \cite{popova2022vulnerability} and reflexes \cite{stiber2022modeling}, somatic expertise and other finely honed physical skills \cite{brymer2010risk}, as well as our ability to put our rational knowledge concerning safety into practice \cite{damasio2006descartes}.

We present an artist-led inquiry into how two professional dancers negotiated safety when engaging in contact dance improvisations with robot arms. This work unfolded over five three-day long workshops held over the course of 18 months. During these workshops, the dancers encountered the robots, gradually gained confidence in employing bodily skills to physically manipulate them, and ultimately learned how to appropriately balance the relationship between aesthetic expression and safety.

Multidisciplinary reflections from the perspectives of HRI, dance and soma design, reveal safety to be a somatic matter---one that holistically engages the mind-body. We propose that HRI needs to expand its current focus on reasoning about safety (anticipation, planning, mitigation and avoidance), to also recognise the vital role of (often pre-conscious) bodily sensations, feelings and skills, in safely negotiating physical contact with robots at close quarters.   


\section{Related Work}




\subsection{Robot and Cobot Safety}\label{sec:rw_safety}

Industrial robots are often placed in guarded areas or cages that prevent direct human-robot contact~\cite{Murashov:2016:Safety,Cheon:2022:Social}. In contrast, collaborative robots (cobots) are typically uncaged, allowing them to work in close proximity with humans~\cite{Michaelis:2022:Uncaged}, requiring alternative approaches to ensuring safety. One approach is to introduce soft boundaries such as \textit{‘light curtains’} (i.e., proximity sensor based safety zones)~\cite{Cheon:2022:BoundedCollaboration} or to limit their speed and force~\cite{Michaelis:2022:Uncaged,Zanchettin:2016:Safety}.
Another is to enable them to anticipate human behaviour, for example, employing multi-modal data (e.g., pose recognition, gaze, video-recordings, and depth-measurements) to model and predict human-motion ~\cite{Yasar:2024:PoseTron}, alongside  dynamic path planning to avoid or mitigate collisions~\cite{patel2023robot,Schirmer:2024:PathPlanning,Heiko:2024:RobotPlanner,Yasar:2024:PoseTron}. This may be supported by a variety of sensors, from overhead cameras to pressure sensitive floor mats~\cite{Michalos:2015:RobotSafety} which may then be used to introduce virtual walls that trigger reactions such as slowing down or stopping when breached \cite{Cheon:2022:BoundedCollaboration}. 
Malik and Bilberg's reference model for HRI proposes four safety layers:
safety monitored stop, hand guiding, speed and separation monitoring, and power and force limiting~\cite{malik2019developing}.

However, despite such efforts, as robots enter the messy human world of everyday life \cite{dobrosovestnova2024ethnography}, so collisions and bodily contact with humans will become unavoidable. Indeed, appropriate social touch will become necessary and even desirable in applications such as rehabilitation \cite{bessler2021safety}, administering care \cite{vallor2020carebots, parviainen2018robots}, and playing and dancing with robots \cite{winfield2022ethical}.
In these situations, we need to move beyond predicting and avoiding contact, to instead consider how to actively embrace it, a stance that requires a radically different approach towards safety.


\subsection{Dancing with Robots}

We focus on dancing with robots as providing a vehicle for creatively exploring the boundaries of human-robot interaction, especially how to safely negotiate bodily contact. While, at first glance, dance might appear to be a niche application for robotics, there is a long history of research into robot dance dating back to the first HRI conference in 2006. At that time, Tanaka et al.~\cite{Tanaka:2006:DancingWithQRIO} focused on how dance could generate knowledge about interaction with robots in uncontrolled and chaotic conditions as we encounter in daily life. They reported that dancing with---or close to---a robot partner might be a desirable activity arising from intrinsic motivation. Similar findings were subsequently reported for non-humanoid robots~\cite{Michalowski:2007:DancingWithKeepon}.  

Rogel et al.~\cite{Rogel:2022:DancingWithPanda} investigated how the movements of professional dancers could help teach a non-humanoid robot to move in a life-like way~\cite{Bartneck2009}, while Frijns et al. considered a humanoid robot mirroring human dance movements~\cite{frijns2024programming}. Jochum and Derks~\cite{Jochum:2019:Dance},  investigated how professional dancers interacted with a non-anthropomorphic mobile robot, while Cuan documented an extensive choreographic exploration of dancing with various robots~\cite{cuan2021dances}, and Gemeinboeck and Saunders
explored feminist relational and participatory perspectives on HRI through dance~\cite{gemeinboeck2023dancing}.
Recent research has also begun to investigate mobile robots as dance partners~\cite{Jochum:2019:Dance,Urann:2020:ChairBot,Knight:2017:ChairBot}.

However, while previous research has established the aesthetic possibilities and wider benefits of dancing with robots, including generating insights for HRI, the dominant focus has been on dancing \textit{near} robots, i.e., without the explicit encouragement of physical contact between human- and robotic partners. We regard this as a research gap, as bodily contact is essential to many forms of dancing. It also inspires us to consider what more we might learn about human interaction with robots, especially with regard to safety, by engaging in more intimate and close quarters dance involving bodily contact.  




\subsection{Somatic Perspectives}
Our focus on bodily contact leads us to adopt a particular \textit{somatic} perspective on interaction. The physical embodiment of computation has been a longstanding concern for both AI~\cite{brooks1991intelligence} and HCI~\cite{dourish2001action} and is a central focus for HRI given its focus on the design of robot bodies. Recently, this focus has broadened to consider the human bodily experience of digital technologies---how can people experience computers through their bodies and how, in turn, might computers transform human bodily experience? In short, how can we put the human body at the centre of interaction design?

Soma design has emerged as a new body-focused methodology to address these questions~\cite{hook2018designing,hook2021unpacking}. This design stance adopts a holistic and pragmatic approach to developing the aesthetic experience of the mind-body, by focusing on how our tacit bodily experiences and practices--- particularly the expertise of professional somatic practitioners \cite{schiphorst2011self}---can inform the design of new technologies. While this perspective has been readily applied to areas such as wellbeing~\cite{hook2019soma} and artistic performance~\cite{avila2020soma}, recent work has explored how it might be applied to the design of physical machines, robots, and drones. La Delfa \& Garrett et al. contribute design concepts that enable more embodied engagement with machines, supporting people in learning about their capabilities and limitations, and allowing them to put technologies to use in different ways \cite{la2024articulating}. Similarly, Sondoqah et al., through a two-day competitive hackathon, demonstrated how somatic engagement with aerial drones can support technical learning and programming \cite{sondoqah2024shaping}. 
LaViers et al. explored how a combination of choreographic and somatic methods might lead to more expressive robot systems\cite{laviers2018choreographic}.
Finally, somatic perspectives can augment our ethical understanding---including safety---of a technology, by framing and considering interactions in terms of how technologies engage the body and the ways in which we act on our ethical understanding of a technology \cite{garrett2023felt}.

To summarise, our review of related work reveals that: we need to reappraise safety as robots increasingly come into close physical contact with humans; dancing with robots provides a powerful vehicle for doing this, as well as being an interesting application in its own right; and we might usefully draw on HCI's recent ‘somatic turn’ \cite{loke2018somatic} to adopt a more holistic and embodied perspective on safety.

\section{Methodology}
We engaged in a process of artistic experimentation in which dancers assessed the potential of robots to inspire novel and aesthetically interesting forms of dance, while also exploring practical concerns including safety, as a first step towards developing a future performance.  
Our approach was exploratory and open-ended; specific research questions were not identified in advance, but rather new knowledge emerged from post hoc reflection on artistic practice \cite{benford2013performance}. 
Similar to Embodied Design Improvisation which elicits tacit knowledge through various methods including ‘domain expert improvisation’~\cite{ sirkin2014using}, our process involved a series of contact improvisation exercises that drew on techniques familiar to both dance and soma design.  
We assembled a multidisciplinary team of four professional dancers and eight researchers with backgrounds spanning HRI, HCI (artist-led methods and soma design) and engineering. 

This work was conducted as part of a project entitled Embodied Trust in TAS: Robots, Dance, Different Bodies. The intention was to build on our previous work with dancers with disabilities, some of whom occasionally wore prostheses when dancing. We had anticipated that their lived experience might yield valuable insights into matters of risk, safety and bodily engagement with robots. 

This team met for five three-day long practical dance workshops over the course of 15 months. The first workshop introduced the team to dance improvisation methods and various industrial, telepresence and quadruped robots. The second involved initial exercises to assess the aesthetic potential of the robots, alighting on  Panda Emika arms as being especially interesting due to the qualities of their movement.
In the third and fourth, the dancers improvised using two connected Franka arms~\cite{haddadin2024franka} that mirrored each others’ movements, taking turns to control and respond to the robots, both live and through the ability to record and replay their movements. 

Dancer improvisations and subsequent group discussions were recorded from multiple perspectives, including a 360-degree view from above. A team member edited the videos to extract 60 examples of dancer-robot interactions. Our team of dancers and researchers reviewed them together, collectively choosing 13 as being of interest for a deeper analysis, while noting emerging themes, including safety. Given space constraints in this paper, three team members selected three examples that best illustrated these themes, transcribed them, and distributed them back to the wider team for comment.

Ethical approval was granted prior to the workshops by Coventry University (reference P151855). Furthermore, following ethical recommendations from previous artist-led~\cite{benford2015ethical} and soma design projects~\cite{garrett2023felt}, we engaged in a processual ethics process by embedding repeated ethical reflections into our research process, frequently revisiting ethical issues, including safety, as we refined our understanding and practices, before presenting ethical reflections to the wider community in this article.

\section{Learning to dance safely with robots}


We present our findings in four parts: a summary of early discussions and explorations that informed protocols, environmental adaptions and supporting roles to enable safe dancing with robots; followed by three examples of dance improvisations that reveal how dancers became familiar with the feel of the robots, how they were subsequently able to dance in aesthetically interesting ways with them, and finally, how an emergency stop was invoked on an occasion when a robot moved in an unexpected way.

\subsection{Establishing safety protocols, environment \& roles}

Our first two workshops focused on encountering various robots, initially with minimal contact, to decide which ones would make interesting dance partners, leading to the decision that the Franka arms had a particular aesthetic appeal to our dancers. Safety briefings revisited manufacturers’ guidelines alongside the usual operating procedures of the robot lab that was hosting us, in the light of the dancers wanting close physical contact with robots. We clarified where and where not to touch the robots and the need to avoid being trapped in joints or in spaces where the arm might bend. We established a technical set-up in which two Franka arms were connected so that (i) one could mirror the others’ movements in real-time and (ii) one robot's movements could be recorded and played back on itself or on the other.
Specifically, Robot 1 could be manipulated by a dancer (the Manipulator). The actuator positions/rotations were recorded and sent to Robot 2, which executed and thus mirrored the motion, thereby functioning as a dance partner to the second dancer (Responder).

We arranged flexible soft furniture to create platforms for the dancers. 
We established protocols governing how to start, shut down and reset the robots, for clearly communicating readiness and the start and end of improvisations, and for when and how to deploy the red emergency stop buttons in case of problems.
These were supported by key operational roles (see Figure \ref{fig:teaserfig}A):
\begin{itemize}
\item Two dancers took on (and swapped) the roles of \textbf{Manipulator}, controlling one robot by holding the Guiding mode buttons at the end of arm and moving it, and \textbf{Responder}, responding to the mirrored movements on the second robot.
\item Two team members were responsible for the \textbf{Manipulator emergency stop button} and \textbf{Responder emergency stop button}. This involved closely monitoring the dancers and robots, and pressing the emergency buttons in case of trouble.
\item A \textbf{Software controller}, sat a a nearby PC, was responsible for operating the control software while a \textbf{Robot wrangler} physical manipulated the robot, for example repositioning it between exercises.
\item An \textbf{Artistic Director} directed the dancers as to the nature of each exercise, when to begin and end, and chose and started the accompanying music.
\end{itemize}

It is important to highlight that, as experienced professionals, our dancers routinely negotiated risks of collisions and falls arising from bodily contact and were highly skilled and experienced in managing these risks, though lacked knowledge of and experience with our robots. We discussed how negotiating an element of risk can be important to the practice of dance, how such skills were gained through practical experience of ‘feeling’ the equipment. We also reflected on the importance of avoiding being overly protective towards the dancers given their disabilities---in order to avoid paternalistic attitudes towards our colleagues~\cite{spiel2020nothing}---in short, we treated the dancers as highly skilled professionals who were learning to interact with a new piece of equipment.



\subsection{Discovering the robots’ limits}

Early exercises with the Franka arms involved the dancers experimenting with how they might best grasp and move them. The Manipulator would typically hold the robot with both hands, sometimes pressing against it with their upper limbs, in order to maximise control while trying to physically feel its response, and often testing the limits of its movement, speed and extension. In turn, the Responder would also sometimes hold it with their hands in order to better feel its movement. However, in doing so, the dancers frequently encountered what they called its ‘freeze’ response; a sudden locking of the arm when its joint positions, motion velocities, accelerations, or torque levels exceeded manufacturer’s preset limits when in this ‘guiding mode’. This brought the improvisation to a premature end as in the following  example (see Video 1).

\begin{figure}[h!]
\begin{tcolorbox}[colback=white, colframe=black, title=Testing the limits \textit{(Video 1, length 2:27 min)}]
{\normalsize
\begin{myquote}


    \texttt{00:31} Holding the robot end effector high above their head, the Manipulator drives the robot's movement in a wide sweeping, seemingly organic, motion. Her intuitive manipulation, completely devoid of reliance on visual cues, as she keeps her gaze fixed on the Responder.

    \texttt{01:08} The Responder, throughout the collaborative dance, has her eyes fixated on the end effector of her robotic dance partner as if attempting to read the robot's movement in a human-like way. She frequently grabs and touches the robot, seemingly improving her felt, bodily connection with it.

    \texttt{01:55} With increasing speed, The Manipulator moves the robot in a back-and-froth reaching motion, resulting in visible enjoyment (\texttt{2:13}). Firmly holding their right hand on the robot's end effector, and their left hand near the sixth joint, they move the robot in a wave-like organic pattern, testing the limits of its capabilities.

    \texttt{02:15} With the increase in robot motion, The Responder moves her entire body---foot to raised hands (see Figure~\ref{fig:teaserfig}A)---dancing with the robot while fixating her gaze on it as if dancing with a human.

    \texttt{02:18} Suddenly, the robot freezes. It has reached its limits. The cause for the perceived breakdown---the manipulator's ever accelerating motions. The Responder's gaze, for the first time, switches to The Manipulator, clearly indicating that she knows that `something' is not as was intended.

    \texttt{02:18} The Manipulator's facial expression clearly shows their realisation that the dance is over. Not by their own choice, but by the limitations of the robot's ability to follow the increasing speed of movement. They briefly stroke the robot as they physically disengage.
\end{myquote}
}
\end{tcolorbox}
    \captionof{vignette}{[from workshop 3, 6th December 2023] The Manipulator explores the limits of the robots freezing behaviour through physical experimentation.}
    \label{vig:NewYork}
\end{figure}

This example illustrates how a safety mechanism, developed to ensure that the robot is used within its operating parameters, creates a source of frustration. The dancers frequently encountered this feature and tried various strategies to better understand, predict and avoid it, including discussing the robot’s behaviour with the robot wrangler on the team, and repeatedly wiggling it back and forth (as we see here), in order to seek out the limit (much as they might do with a responsive human partner). However, neither strategy---talking through the problem or physically trying to sense the ‘freeze point’---proved fully satisfactory. Ultimately, the dancers had to learn to manipulate carefully and rely on slower movement. 

It became clear to the observers in the room, that the Manipulator was experimenting with the robot in order to know its capabilities: \textit{``...the wiggling and wriggling is really interesting. It does feels like [anon], you’re pushing and testing the limits [...]''}. The Responder, while focusing on her own robot partner, noticed that the Manipulator and their robot were expressing seemingly orchestrated movement. She stated that \textit{``It seemed like you were more in the kind of choreographic space in a way. [...]] when you are in that choreographic space, you know where the limits can be tested and then it just gives up [robot stops], you know, it is giving up.''} 






\subsection{Performing vulnerability}\label{sec:performingVul}
Our second example (Video 2), occurring two months later, sees the dancers (now swapped in their roles of Manipulator and Responder) being increasingly comfortable with the robots and so able to more creative explore their aesthetic potential. One aesthetic that emerged was to ‘perform vulnerability’, with the dancers \textit{apparently} willing to take risks as they seemed to expose their bodies to possible collisions. Throughout the sequence below, the Manipulator has her eyes closed, focusing on controlling her robot through her bodily senses and being freed up from having to worry about the Responder, who in turn, then appears to take a number of risks such as turning their back on the robot as it moves.

\begin{figure}[h]
\begin{tcolorbox}[colback=white, colframe=black, title=Taking risks \textit{(Video 2, length 1:43 min)}]
{\normalsize
\begin{myquote}
    \texttt{00:00} Responder is sitting down, slightly tilting their head to the right and looking attentively at the arm which is being as controlled by Manipulator with her eyes shut. 
    
    \texttt{00:10} As the arm raises upwards, Responder bends forward underneath it, exposing her back to the robot. 
    
    \texttt{00:33} 
    Responder twists around to face the robot, exposing her torso with her arms wide open, closely following its movements  to keep a small separation between them. 
    
    \texttt{00:42} Responder stands up, raising their left arm close to the robot's gripper while swinging their right arm closely underneath the arm. 
    
    \texttt{00:45} Responder appears to wobble on her prostheses and, as the robot arm arches downwards, puts their forearm on top of it, slowly brushing their hand along the length of the arm robot while turning away from it until just fingertips are touching, before stepping away. 

   \texttt{01:02} Responder faces the robot and physically reengages by touching it from underneath with their right arm, before twisting to expose their face to it from below.
    
    \texttt{01:15} Responder crouches underneath the robot arm, when suddenly Manipulator swings it to its left and down, making slight contact with Responder's back who, feeling it, quickly ducks down. Responder Button who is watching closely, presents a clearly distressed grin and tenses her shoulders when observing this collision but does not press the emergency stop button. 
    
    \texttt{01:31} Towards the end of the dance, Responder stands up while keeping close contact with the robot, continuing to swing their arm across it, interlacing with it and once again turning their back on it.
\end{myquote}
}
\end{tcolorbox}
    \captionof{vignette}{[from workshop 4, 6th February 2024] Once familiar with the robot, the dancers are able to take risks as part of an aesthetic of performing vulnerability.}
       \label{vig:eyes}
\end{figure}


Early in this improvisation, the Responder can be observed looking attentively at the robot while conducting their initial movements at a distance. As the dance unfolds, their movements gradually became looser, more expansive, and more proximate to the robot. Throughout the performance, they explore different contacts in close contact with the robot, sometimes brushing it with their hands or arms. As the improvisation progresses further, they become more adventurous, eventually exposing their back to the robot, by getting underneath it. This resulted in near misses and a brief collision of the gripper with their back as the Responder Button-holder looks on with a worried expression, their hand hovering over the red button, but without pressing it (see Figure~\ref{fig:teaserfig}C). 

In the subsequent discussion, both Manipulator and Responder commented that they found this improvisation to be particularly rewarding aesthetically. The Responder observed how they had intentionally risked turning their back on the robot, but had sensed how it was moving and knew how to drop to the ground to avoid it: \textit{``It was one of the nicest moments I’ve had with [the robot.] I felt ‘okay, I have to go to my knee.’ It’s like when you’re dancing with a human and the lead is so clear like ‘okay, you want me to go there. So I will go there.''}
In turn, the Responder Button-holder observed how they felt uncomfortable and had nearly pressed the button, but ultimately felt able to trust to the Responder’s judgement and skill: \textit{``I felt that [the Responder] had done this intentionally. And throughout, I feel [their] style of dancing with the robot has been purposefully provocative and subversive.''} adding \textit{``So I was like, this is just [anon] dancing. And I’m uncomfortable... But then should I act on that, should I make the decision to shut it down?''}




\subsection{A problem occurs}\label{sec:problemOccurs} 

While the overwhelming majority of interactions proceeded without incident, there was one case in which the arm moved in way that was unexpected, causing surprise, and could have been potentially dangerous, which led to an emergency button being pressed.
This arose when experimenting with the capability to replay recorded movements. 
By mistake, a recording captured from one robot was played back on the other robot. 
While the robots might appear to be identical, they are operating in different environments, the manipulator robot is in a tighter space featuring raised seating with an enclosing back, while its mirrored responder robot is in a more open space that affords greater range of movement, especially downwards towards the floor. 

Analysis of a forty second long sequence below (Video 3) reveals how participants gradually become aware that the robot is moving in an unexpected way, communicating this tacitly through a series of glances, movements and concerned noises, before the red button gets pressed. 

\begin{figure}[h]
\begin{tcolorbox}[colback=white, colframe=black, title=An emergency stop \textit{(Video 3, length 40 sec)}]
{\normalsize
\begin{myquote}
    \texttt{00:00} A dancer is relaxing and leaning back on the cushions with their arms outstretched, and their prostheses removed, watching the nearby robot while its movement is being tested prior to a dance. The red button operator is watching the robot intently, with her hand poised over her red button. 
    
    \texttt{00:12} The button operator shifts her gaze to be able to track the end of the robot as it moves lower, while the software controller looks away from his console towards the robot. 
    
    \texttt{00:20} As the robot continues to move, the dancer shifts both their arms inwards (perhaps ready to defend against the robot), and glances at the button operator. The software operator visibly tracks the robot closely, makes a concerned noise, and stands up from his chair.  
    
    \texttt{00:27} At this point the robot end effector has reached its highest point and changes direction quickly moving towards the dancer's legs. 
    
    \texttt{00:29} The end effector of the arm moves close to the dancer's left thigh, causing the button operator to press her red button. The power shuts off and the robot freezes with a loud clunk. An audience member says ``ooh'', button operator says ``sorry'', and Software controller says something unintelligible. Opposite them the second dancer and their button operator briefly glance over apparently unconcerned, while their robot continues moving.
    
    \texttt{00:34} A different audience member remarks ``very close'', as if approving Sarah's decision to press the button.
\end{myquote}
}
\end{tcolorbox}
    \captionof{vignette}{[from workshop 4, 6th February 2024] Growing awareness in the room of strange robot behaviour leads to the `Red Button' being pressed}
    \label{vig:RedBtn}
\end{figure}

A lengthy debriefing shed light on this incident. First, in contrast to the previous examples, the dancer was not readied for action, but rather was leaning back in a relaxed position. As they observed:
\textit{``you’re not in that position as a performer, you’re not here choosing to take risks in the same way. You’re sitting back looking at the robot like the rest of us. So you’re not knowingly exposed in the same way. Now the robot’s doing something that no one’s predicting''}, further adding \textit{``I guess it’s like what stage are you in and what stage are you out? Which we do talk about in dance studios a lot.''} The inward movement of their arms [00:20] signals a change in readiness in response to growing concern about the robot.
 
We highlight how a growing awareness that something is wrong was somatically enacted among the various parties through a series of glances, movements and noises, but notably without speaking, over a period of ten seconds, providing a series of visible cues of an impending problem. This may be in part because those concerned had different viewpoints and perhaps different tolerances: \textit{``So there’s three potential different peoples that will all necessarily have different stopping points.''} 

The red button operator reported being unsure of whether she had pressed the button properly, suggesting a need to practice in advance in order to get the feel of it:\textit{``I was thinking, my God, I haven’t pressed it properly.''}, adding \textit{``I was more worried about like if I made the decision to shut down the robot when [anon] is having a great time doing their cool thing.''} The dancer reflected on how one also needs to learn the consequences of being hit by the robot: \textit{``So I don’t think I knew what it felt like to be hit by a robot. I had quite a bit of practice in it, but we had both [found it] really hard to see the consequences of what might happen.''} This led to a discussion of who should work the button, an expert in robotics or an expert in dance?
Finally, this example reveals how seemingly identical robots are not identical in practice due to their context, in this case the presence or absence of the cushions. Indeed, the very features that were added to improve safety, i.e., the soft furniture behind the dancer (see Figure~\ref{fig:teaserfig}B), may have exacerbated the problem by restricting their ability to escape.

\section{Discussions \& Reflections}


Our experience of improvising contact with robots has revealed the vital role of the body in safety. Touching and grasping robots, as well as feeling, moving, and tacitly communicating a growing sense of unease, all relied on bodily interaction. The dancers drew on carefully honed bodily skills, alongside growing knowledge of how the robots ‘felt’, acquired through hands-on practice. This leads us to define \textit{somatic safety as a holistic approach in which safety is learned and enacted through the body as well as being reasoned about in the mind; that recognises the body to be a site of capability, not only vulnerability; and that treats safety as an ongoing negotiation between humans and robots through contact.} 

Previous research has largely focused on what we call ‘safety in the mind’ in which, humans anticipate and plan safety before physically engaging robots through guidelines, regulations, plans, and risk assessments~\cite{zacharaki2020safety}, while in turn, robots are programmed to anticipate and avoid contact through human behaviour analysis and motion planning~\cite{lasota2017survey}. Where the body is considered, it is often treated as vulnerable and subject to possible injury. This may be sensible in industrial or vehicular contexts where robots are dangerous and there are realistic possibilities of separating them from humans. However, it will be insufficient as robots spread into messy everyday contexts~\cite{la2024articulating}, where bodily contact will become unavoidable, necessary, or even desired, as robots are increasingly used for therapy, care, play, and even dance.

Over time, humans acquire a repertoire of bodily skills. From basic learning through rough-and-tumble childhood play, to acquiring sophisticated bodily awareness through sport and dance, humans learn how to safely engage the world through physical experience which underpins our reasoning about what we can do. Somatic safety recognises the importance of this, and calls for harnessing such skills, as part of our future interactions with robots. Disciplines as diverse as neuroscience~\cite{damasio2006descartes} and philosophy~\cite{johnson2007meaning}, recognise that human actions and reactions are deeply embodied, spanning autonomic systems, reflexes and emotions. Ensuring safe interaction with robots in real-time, needs to accommodate the preconscious nature of our actions and reactions---how we might ‘feel’ before ‘thinking’, which shapes processes of thinking, planning, and reflection (and vice-versa). While previous research has considered how such human responses can be used to predict errors~\cite{stiber2022modeling}, we view them as being inherent to somatic safety as a skill that can be developed---if we can design to support it. Somatic safety is not a matter of having bodily responses nor monitoring bodily responses to robots. Rather, it considers how the physical bodies of robots and the modalities through which we interact with them, can help people to develop deeper understandings of these technologies through their bodies. This said, we emphasise the importance of also thinking about potential risks and best practice before first encountering an unfamiliar robot if possible, as was the case in our own research. In short, somatic safety entwines physically enacting and cognitively reasoning about safety. Having introduced the broad idea of somatic safety using the domain of dance, we now draw out three key implications for HRI research.

\subsection{Learning safety through bodily experience}
Somatic safety needs to be learned and adapted to new situations, such as encountering a new and unknown robot. It is vital to gain repeated direct bodily experience~\cite{hook2018designing,hook2021unpacking} of manipulating robots, and even of the consequences of being hit by them (within safe limits) in order to fine tune one’s somatic skills. La Delfa \& Garrett et al. have shown that it is possible to understand the capabilities and limitations of a machine through somatic engagement~\cite{la2024articulating}. Similarly, Sondoqah et al. demonstrate that somatic engagement with nano-drones, particularly crashes, can augment technical knowledge about a system~\cite{sondoqah2024shaping}. This suggests that a holistic perspective is a promising approach to support successful HRI, and we extend this perspective towards safety. 

Designing robots to better support our learning can help us develop somatic safety skills. In turn, we can put those skills into practice; knowing where and how to touch a robot, what it is capable of and what not to do (even without technical expertise), and even where the boundaries of risk and safety may be negotiated. Somatic safety is intended to augment, rather than replace, other approaches to safety. We combined somatic perspectives with traditional safety approaches such as safety briefings. Through repeated reflections, we were able to identify spaces where existing approaches could not account for the complexity of the interaction and may have even posed a hindrance. 

The dancers’ lived experience of disability was invaluable as it is important to respect people’s existing somatic expertise, treating our dancers as highly skilled practitioners, rather than vulnerable individuals in need of paternalistic protection \cite{bennett2019promise}. However, even highly skilled practitioners will be unfamiliar with a new equipment and will need to be trained how to use it safely. We argue that this training needs to be practical as well as theoretical, and suggest that HRI can pay attention to the tight iterations between bodily practice and reflection as part of a holistic somatic approach to safety.

\subsection{Making robots touchable and graspable}
Robots could also be designed to support developing somatic skills through physical interactions and feedback mechanisms. Touching and grasping was necessary to manipulate the robots. However, we suggest it also led to safer interactions. Grasping, either with limbs or through other mechanisms, enables humans to engage directly through haptic, tactile and kinaesthetic senses. Taking hold better enables us to constrain something that we are trying to control, offering a platform against which we can push if we need to quickly launch ourselves away. In short, as the dancers showed, if you want to control something, you are often better to take hold of it so you can fully employ your embodied senses and skills. 
However, it is then important to know where and how to grasp robots. Our extensive safety briefings and reflections pointed out how some parts of the robot were more safely touchable while others were not (e.g., joints). While the normal mode of guiding these robots is through the guiding-mode buttons, the dancers' interactions demonstrated how more controlled movement could be achieved through additional physical contact at other points.
However, the robots' various surfaces are not clearly designed to afford or discourage grasping. Could robots be given small handles, finger holds, bevels, and tethers to afford safe and effective grasping, rather than having smooth and undifferentiated surfaces as with many industrial robots (provided these could be designed to avoid causing additional complications such as the robot colliding with itself)? We might also extend soft skins for cobots (e.g, Airskin\footnote{www.airskin.io/airskin}) to be capable of a greater range of sensing and physical actuation.

Through interactions between robots and dancers, we saw how the challenges of physically moving the robot extended to the control software. For instance, while mechanisms such as freezing are intended to be a safety mechanism as discussed in Section~\ref{sec:rw_safety}, in practice they reduced our dancers’ abilities to employ their bodily skills to sense when they were approaching the limits of safety. Whereas contact improvisation with a human involves feeling and learning the limits of the other person through repeated movements and a sensitivity to their bodily response, this was not possible with the robots. Rather, the sudden stop experienced during the movements when approaching the Franka Control Interface limits defined by the manufacturer might actually have been a hindrance to developing an intuitive/somatic learning of these limits. The implication is that robots, like people, should have soft boundaries at the limits of safety, signalling discomfort through resistance and other cues, before they reach the point of freezing. We argue this could support people in developing somatic skills to interact with robots, rather than preventing the possibility of developing skills by constraining the interaction.


\subsection{Visibly enacting safety through the body} 
Our explorations revealed two ways in which safety is somatically enacted for others in a social situation. The first was as an overt performance, in our case, a dance performed to a watching audience. Our dancers carefully invoked a palpable sense of vulnerability and risk in the audience when they danced close to the robot, especially when they exposed more vulnerable parts of their bodies. They were able to present such \textit{apparent} risk-taking to the audience, due to their increasing familiarity with the robot and somatic safety skills that allowed them to negotiate the boundaries of interaction. One implication is that ‘spotters’ in charge of ‘red buttons’ should be capable of distinguishing carefully performed risk-taking (in e.g., ‘Performing vulnerability’, see Section~\ref{sec:performingVul}) from unwitting risks of which dancers are less aware (in e.g., ‘A problem occurs’, see Section~\ref{sec:problemOccurs}). In our process, this relied on familiarity with both individual dancers (somatically understanding dancing style) and robots (somatic understanding different physical environments).

Second, there was a more implicit performativity of safety that became apparent in the moment of pressing the red button. Here, others were able to witness people’s bodily responses to an emerging safety concern, including subtle shifts in posture signalling, readiness to react, or movements and glances that signalled closer attention to the robot and growing concern. Research has already looked at how human responses can be used to predict errors; however, this might rely on systems such as facial recognition~\cite{stiber2022modeling}. We see potential here for different forms of embodied or somatic communication between humans and robots (e.g.,~\cite{la2024train}), which can be used to indicate or develop an intuition for when something is going wrong, without the need for invasive monitoring. There is already a significant body of work on socially responsive robots~\cite{henschel2021makes}, but such robots rely on a very narrow and normative range of socially meaningful behaviours such as gaze and gesture. We also see potential to develop a wider range of ways in which robots can in their turn physically ‘perform’ safety or vulnerability that are perceptible to humans and indicate the emergence of a risk or limitation of the technology.

\section{Conclusion}
As robots inevitably and for some, desirably, come into evermore intimate physical contact with humans, so we need to change our approach to safety. We can no longer treat safety as being a matter to reason about at a distance, but rather need to adopt a somatic perspective in which safety is felt, learned and enacted through ongoing bodily contact, interleaved with frequent reflection. We see somatic safety as broadening our approaches to ethics in HRI. Often, we treat ethics as a matter of procedure: of a set of rigid guidelines to follow. However, somatic safety approaches ethical HRI as a ongoing process of ‘processual ethics’ through which we continually learn about the best ways to interact through bodily practice~\cite{garrett2023felt}. A further ethical consideration is to empower people with varying abilities to engage robots on their own terms rather than treating them as being vulnerable and risking being overly protective. Future work might explore the application of somatic safety beyond artistic applications such as dance. Robotics within care and rehabilitation would require such considerations, though ultimately, somatic safety will become even more widely relevant as robots continue to spread into the messy everyday human world.

\section*{Acknowledgments}
This work was supported by the UKRI through the Turing AI World Leading Researcher Fellowship: Somabotics: Creatively Embodying Artificial Intelligence, and by the Engineering and Physical Sciences Research Council [grant number EP/V00784X/1] UKRI Trustworthy Autonomous Systems Hub and Responsible AI UK [grant number EP/Y009800/1]. We are grateful to Dr Joseph Bolarinwa, Bristol Robotics Laboratory,  for the software to connect the two cobots.


\bibliographystyle{IEEEtran}
\IEEEtriggeratref{37}
\IEEEtriggercmd{\enlargethispage{-4.7in}}
\bibliography{MAIN}


\end{document}